\documentstyle[pre,aps,preprint,epsfig]{revtex}
\tightenlines
\input{epsf.tex}

\begin{document}

\title{Short-range plasma model for intermediate spectral statistics}
\author{Eug\`ene Bogomolny$^1$, Ulrich Gerland$^{2}$ and 
        Charles Schmit$^1$}
\address{$^1$Laboratoire de Physique Th\'eorique et Mod\`eles 
         Statistiques\thanks{Unit\'e de Recherche de l'Universit\'e Paris XI
	 et du CNRS (UMR 8626)}, Univ. Paris-Sud, 91405 Orsay Cedex, France}
\address{$^2$Phys. Dep., UCSD, La Jolla, CA 92093-0319, USA}

\date{\today}

\maketitle


\begin{abstract}
We propose a plasma model for spectral statistics displaying level
repulsion without long-range spectral rigidity, i.e. statistics
intermediate between random matrix and Poisson statistics similar to
the ones found numerically at the critical point of the Anderson
metal-insulator transition in disordered systems and in certain
dynamical systems.
The model emerges from Dysons one-dimensional gas corresponding to the
eigenvalue distribution of the classical random matrix ensembles by
restricting the logarithmic pair interaction to a finite number $k$ of
nearest neighbors.
We calculate analytically the spacing distributions and the two-level
statistics. In particular we show that the number variance has the
asymptotic form $\Sigma^2(L)\sim\chi L$ for large $L$ and
the nearest-neighbor distribution decreases exponentially when 
$s\rightarrow \infty$, $P(s)\sim\exp (-\Lambda s)$  with
$\Lambda=1/\chi=k\beta+1$, where $\beta$ is the inverse temperature of the
gas ($\beta=$1, 2 and 4 for the orthogonal, unitary and symplectic
symmetry class respectively). In the simplest case of $k=\beta=1$, the model 
leads to the so-called Semi-Poisson statistics characterized by particular
simple correlation functions e.g.  $P(s)=4s\exp(-2s)$.
Furthermore we investigate the spectral statistics of several
pseudointegrable quantum billiards numerically and compare them 
to the Semi-Poisson statistics. 
\end{abstract}

\pagebreak

\section{Introduction}
Since the advent of random matrix theory there has been
considerable interest in the statistical analysis of spectra
\cite{porter,mehta,bohigas89,guhr98a}.  
Two diametrically opposed statistical distributions have been found 
to be of universal relevance: 
the Poisson distribution, i.e. completely uncorrelated levels, 
and the Wigner-Dyson distributions of random matrix theory. 
The prominent features of these distributions are conveniently 
characterized with the help of spectral observables such as the 
nearest-neighbor spacing distribution $P(s)$ and the number 
variance $\Sigma^2(L)$ (the variance of the number of levels in an 
energy interval of length $L$ in the unfolded spectrum). 
The former stresses the correlations on a short scale, while the 
latter measures the stiffness of the spectrum, i.e. long-range 
spectral correlations.
For the standard random matrix ensembles (orthogonal, unitary and
symplectic symmetry labeled by $\beta=$1, 2 and 4 respectively) the
spacing distribution is approximately 
given by the Wigner surmise and the number variance increases
only logarithmically with large $L$,
\begin{eqnarray}
  \label{Pwigner}
  P(s) &=& a_\beta\,s^\beta\,\exp(-c_\beta\,s^2)
  \nonumber \\
  \Sigma^2(L) &\sim& \frac{2}{\beta\pi^2}\log(L)\qquad(L\rightarrow\infty)
\end{eqnarray}
($a_\beta$ and $c_\beta$ are determined by normalization and by
requiring the mean level spacing to be one). 
Thus the Wigner-Dyson distributions are marked by level repulsion and
long-range spectral rigidity. 
In contrast, for the Poisson distribution one has neither
level repulsion nor spectral rigidity, 
\begin{eqnarray}
  \label{Poisson}
  P(s) &=& \exp(-s) \nonumber \\
  \Sigma^2(L) &=& L \;.
\end{eqnarray}
In the present article we devise a discrete set of statistical
distributions with properties intermediate between those of the 
Poisson and Wigner-Dyson distributions, namely level repulsion 
\begin{equation}
  \label{levrepul}
  P(s)\sim s^\beta \qquad(s\rightarrow 0)
\end{equation}
paired with an exponential decay of the nearest-neighbor spacing 
distribution, 
\begin{equation}
  \label{expdecay}
  P(s)\sim \exp(-\Lambda s)\qquad(s\rightarrow\infty)\;, 
\end{equation}
and a linear asymptotic of the number variance, 
\begin{equation}
  \label{asymnv}
  \Sigma^2(L)\sim\chi L \qquad(L\rightarrow\infty)\;.
\end{equation}
Our distributions arise as the Gibbs distribution of a 
one-dimensional gas of interacting classical particles.
It was noted already by Dyson that the distribution of eigenvalues $E_j$
of random matrices is equivalent to the distribution of particles in a
one-dimensional gas with inverse temperature $\beta$ and the repulsive 
two-body interaction potential $V(x)=-\log|x|$, since the joint 
probability distribution can be written in the form
\begin{equation}
  \label{gibbs}
  P_{N\beta}(E_1,\ldots,E_N)=Z_N^{-1}\, {\rm e}^{-\beta\,
  W(E_1,\ldots,E_N)} \;,
\end{equation}
where the normalization constant $Z_N$ is the canonical partition
function of the gas and 
\begin{equation}
  \label{totenergy}
  W(E_1,\ldots,E_N)=\sum_i U(E_i)+\sum_{i<j}\,V(|E_i-E_j|)
\end{equation}
is the total potential energy.
The one-body potential $U$ serves to confine the $E_j$ to a finite 
stretch of the $E$-axis, e.g. $U(E)\propto E^2$ for the Gaussian 
ensembles. It determines the mean particle (level) density, but
it is irrelevant for the statistical correlations on the scale of the 
mean level spacing (random matrix universality) as long as $U$ is 
steep enough to actually confine the levels \cite{pato}.

Our plasma model of intermediate spectral statistics is defined by the 
same eqs.~(\ref{gibbs}) and (\ref{totenergy}), except that we restrict 
the second sum on the right hand side of eq.~(\ref{totenergy}) to a 
finite number $k$ of nearest neighbors.
It is this screening of the Coulomb interaction between the levels, 
which is the essential ingredient leading to the properties 
eqs.~(\ref{levrepul})-(\ref{asymnv}).

Our first motivation stems from the physics of disordered electronic
systems. A $3$-dimensional disordered sample undergoes 
a phase transition between an insulating and a metallic phase
as a function of the disorder strength (Anderson
metal-insulator transition). In the insulating phase the electron 
eigenfunctions are localized and since non-overlapping eigenfunctions
are uncorrelated, the eigenenergies are Poisson distributed. 
In contrast, in the metallic phase the eigenfunctions extend
homogeneously over the whole sample and overlap strongly which leads
to a Wigner-Dyson distribution of the energy levels. 
Exactly at the transition point the electron eigenfunctions are
extended, but strongly inhomogeneous (multifractal). This leads to 
intermediate spectral statistics, which are believed to be universal,
i.e. independent of the microscopic details of the disordered 
system \cite{shklovskii93a}. 

Fyodorov and Mirlin \cite{fyodorov97a}  calculated the overlap of two critical
eigenstates with slightly different energy and found that the overlap
is one if the energy separation, $s$, between
the eigenfunctions is of the order of  mean level spacing, while for
larger $s$ it decays as a certain power of $s$.
In contrast, for Wigner-Dyson statistics the overlap 
is one for all values of $s$. Hence one may conclude that at the 
critical point of the Anderson metal-insulator transition only 
eigenvalues which are separated by at most a few level spacings
interact strongly.

A second motivation which led us to consider the 
short-range plasma model is the following.
In the context of quantum chaos the statistical analysis of spectra 
plays a major role. In the semiclassical limit the quantum energy 
spectra of systems with integrable classical dynamics generally 
display Poisson statistics \cite{berrytabor}, while the Wigner-Dyson
distribution is associated with fully chaotic classical dynamics
\cite{bohigas84}. But there are systems  which are neither integrable 
nor chaotic. In  \cite{bogomolny99a} we examined numerically a few 
such systems and found that their spectral statistics exhibit  all 
properties (\ref{levrepul}) - (\ref{asymnv})   typical for intermediate
statistics. 
In particular, the simple expression
\begin{equation}
  \label{semifish}
  P(s)=4s\,e^{-2s}
\end{equation}
is an excellent fit to the spacing distribution of a subclass of systems
considered. We further noticed that the short-range plasma model with 
nearest-neighbor interaction ($k=1$) leads precisely to this 
distribution (and we checked that other correlation functions are also well
described by this model).

Another interesting class of plasma models with screened Coulomb 
interaction is the Gaudin model \cite{gaudin66a,forrester93a} defined by 
eqs.~(\ref{gibbs}) and (\ref{totenergy}) with the two-body interaction 
potential
\begin{equation}
  \label{gaudin}
  V(x)=-\frac{1}{2}\log\frac{x^2}{a^2+x^2}\;.
\end{equation}
When $a\rightarrow \infty$ it reduces to the random matrix models, while in
the limit $a\rightarrow 0$ it yields Poisson statistics.
For $\beta=2$ the model is solvable and all correlation functions can
be written in a closed form \cite{gaudin66a,forrester93a}. They obey all
three conditions (\ref{levrepul}) - (\ref{asymnv}) characteristic
for intermediate statistics with the following constants (for $\beta=2$)
\begin{equation}
\chi= \frac{1}{\alpha}(1-e^{-\alpha}),
\label{chi}
\end{equation}
and
\begin{equation}
\Lambda=\frac{1}{\alpha}\int_{0}^{e^{\alpha}-1}\frac{\ln (1+t)}{t}dt.
\label{lambda}
\end{equation}
Here $\alpha=2\pi a \rho$ and $\rho$ is the mean density of the levels.

In general, in any plasma model with 
short-range interaction the nearest-neighbor spacing distribution will 
decay exponentially at large distances. It is the otherwise unusual 
property of long-range interaction in the standard random matrix models 
(which manifests itself in the summation over all pairs of particles in 
eq.~(\ref{totenergy})) that is responsible for the $\exp(-cs^2)$ decay of 
$P(s)$ at large $s$. 

The main drawback of the Gaudin model is that no analytical solution is 
known for $\beta\neq 2$. In contrast, our model can be solved for
arbitrary $\beta$ (and arbitrary potential).

We stress that our one-dimensional gas model is meant as a toy model 
which deserves interest, because it provides a natural discrete 
interpolation between Poisson and Wigner-Dyson statistics, it is 
analytically solvable, and it is physically motivated. 
It does not pretend to furnish an exact description of the critical statistics
of a physical model (e.g. the  Anderson model in the MIT point).
Despite significant theoretical progress (see \cite{fyodorov97a},
\cite{kravtsov}, \cite{kravtsov95}-\cite{tsvelik} and references therein), the
precise form of the critical distribution is not yet known though there
exist arguments (\cite{kravtsov95}-\cite{tsvelik}) that the main difference
between the standard random matrix ensembles and the intermediate statistics
is in the form of the correlation kernel, $K(x,y)$, which determines the correlation
functions in the random matrix theory \cite{mehta}. In the standard ensembles 
\begin{equation}
K(x,y)=\frac{\sin \pi(x-y)}{\pi(x-y)}.
\end{equation}
For intermediate statistics it has been argue that
\begin{equation}
K(x,y)\approx \frac{a\sin \pi(x-y)}{\pi \sinh a (x-y)},
\label{kernel}
\end{equation}
where $a$ is a parameter. When $a\rightarrow 0$ the standard random
ensembles are recoved. Non-zero value of $a$ characterizes the intermediate
statistics. The detailed discussion of this type of critical behavior is
beyond the scope of this paper.

The paper is organized as follows. In the next section we present  
the calculation of the spacing distributions and the number variance 
in the one-dimensional plasma, first for $k=1$, then for $k=2$, and 
finally for arbitrary $k$.
We compare the resulting distributions with our numerical results 
for the pseudointegrable billiards in section~\ref{pseudo}. 
Finally, in the discussion of section~\ref{discussion} we compare the 
short-range plasma model to other existing interpolations between 
Poisson and Wigner-Dyson statistics.
\section{The model}
\label{model}
Instead of a linear one-dimensional gas as in eqs.~(\ref{gibbs}) 
and (\ref{totenergy}), where the levels are confined by a one-body 
potential $U(E)$, we choose a circular geometry. This is convenient, 
since on a circle the levels are automatically confined and it is 
unnecessary to introduce an external potential. The mean level density 
is then constant and unfolding is trivial. At the same time the 
correlations in the unfolded level sequence are the same as in the 
linear geometry (in the limit of a large number of levels), just as 
the Gaussian ensembles of random matrix theory are locally equivalent 
to Dyson's circular ensembles. 
The method of calculation that we apply in this section goes back 
to G\"ursey \cite{gursey50a} for the case of nearest-neighbor 
interaction, and to van~Hove \cite{vanhove50a} for the general case 
(see also ref. \cite{lieb66a}).

\begin{figure}[t]
 \begin{center}
  \leavevmode
  \epsfig{figure=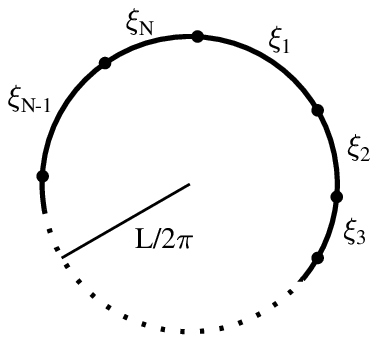}
 \end{center}
 \caption{$N$ particles on a circle of circumference $L$. 
   The positive spacings between 
   consecutive particles measured along the circle are denoted by
   $\xi_j$ with $j=1\ldots N$.}
 \label{circle}
\end{figure}

We consider $N$ particles (representing energy levels) on a circle of
circumference $L$. We denote the positive spacings between
neighboring particles by $\xi_j$ with $j=1\ldots N$ (see
Fig.~\ref{circle}). Hence, 
\begin{displaymath}
  \sum_{j=1}^N \xi_j=L\;.
\end{displaymath}
For convenience we use a periodic index, i.e. $\xi_{j+N}:=\xi_{j}$.
We introduce an interaction between the particles via a repulsive
two-body potential $V(\xi)$ (eventually we will choose 
$V(\xi)=-\log |\xi|$ as in random matrix theory), but we let each
particle interact only with its $k$ nearest neighbors to the left and
to the right $(k=1,2,3,\ldots )$. 

The canonical partition function of this one-di\-men\-sional gas is 
\begin{eqnarray}
  \label{ZN}
  Z_N(L,\beta)&=&\int\limits_0^{\infty}{\rm d}\xi_1\ldots
  \int\limits_0^{\infty}{\rm d}\xi_N\;
  \delta\Big(L-\sum_{i=1}^N\xi_i\Big)\times \nonumber \\
  & &\exp\Big(-\beta\sum_{j=1}^{N}W(\xi_j,\ldots,\xi_{j+k-1})\Big)\;,
\end{eqnarray}
where 
\begin{eqnarray}
  \label{W}
  W(\xi_j,\ldots,\xi_{j+k-1}) &=& V(\xi_j)+V(\xi_j+\xi_{j+1})
  +\nonumber\\ & & \ldots+V(\xi_j+\ldots+\xi_{j+k-1})\;.
\end{eqnarray}
$W(\xi_j,\ldots,\xi_{j+k-1})$ includes the interaction
energy of the particle which is located at the left of $\xi_j$ with
its $k$ nearest {\sl right} neighbors only to avoid double counting. 
Since for the application to spectral statistics $\beta$ is not a free
variable, but takes only the values 1, 2, and 4, we suppress the
$\beta$-dependence of the partition function in the following. For
brevity of notation we define  
\begin{equation}
  \label{fofx}
  f(\xi):=\exp\Big(-\beta\,V(\xi)\Big)\;,
\end{equation}
which for the case of $V(\xi)=\ln |\xi|$ becomes
\begin{equation}
  \label{frm}
  f(\xi )=|\xi |^\beta\;.
\end{equation}
The joint probability distribution of $n$ consecutive spacings 
then takes the form 
\begin{eqnarray}
  \label{jointspace}
  p(\xi_1,\ldots,\xi_n) &=& \frac{1}{Z_N(L)}\,
  \int\limits_0^{\infty}{\rm d}\xi_{n+1}\ldots
  \int\limits_0^{\infty}{\rm d}\xi_N\;
  \delta\Big(L-\sum_{i=1}^N\xi_i\Big)\nonumber \\
  & & \hspace*{-2cm} \times 
  \prod_{j=1}^{N} f(\xi_j)f(\xi_j+\xi_{j+1})\cdots
  f(\xi_j+\ldots+\xi_{j+k-1})\;.
\end{eqnarray}

In the following the variable $s$ denotes distances measured in units
of the mean spacing $\Delta=L/N$.
Our aim is to calculate correlation functions such as the $n$-th
nearest-neighbor spacing distribution $P(n,s)$ 
(the distribution of the distance $s$ between two particles that have
$n$ particles between them) and the two-point correlation function
$R_2(s)$ (the probability of finding any two particles at distance
$s$). The latter is related to the former by summation over $n$,
\begin{equation}
  \label{relR2Pn}
  R_2(s)=\sum_{n=0}^{\infty} P(n,s)\;.
\end{equation}
Using eq.~(\ref{jointspace}) the $n$-th nearest-neighbor
spacing distribution can be expressed as 
\begin{eqnarray}
  \label{nthP}
  P(n,s) &=& \int\limits_0^{\infty}{\rm d}\xi_{1}\ldots
  \int\limits_0^{\infty}{\rm d}\xi_{n+1}\; \delta\Big( 
  s-\frac{1}{\Delta}\sum_{i=1}^{n+1}\xi_i\Big)\times \nonumber\\ & &
  \qquad\qquad\qquad p(\xi_1,\ldots,\xi_{n+1})\;.  
\end{eqnarray}
The number variance can be obtained from the two-point correlation
function with help of the relation
\begin{equation}
  \label{relsigma}
  \Sigma^2(L)=L-2\int\limits_0^L{\rm d}s\,(L-s)(1-R_2(s))\;.
\end{equation}
A convenient way to calculate the asymptotic behavior of the number
variance for large $L$ is to consider the asymptotic expansion of the
Laplace transform of the two-point correlation function, 
\begin{equation}
  \label{LaplaceR2}
  g_2(t)=\int\limits_0^{\infty}{\rm d}s\;R_2(s)\;e^{-ts}\;,
\end{equation}
for small $t$. If 
\begin{equation}
  \label{g2expansion}
  g_2(t)=\frac{1}{t}+\alpha_0+\alpha_1 t +O(t^2) \qquad 
  (t\rightarrow 0)
\end{equation}
then we get from eq.~(\ref{relsigma})
\begin{equation}
  \label{asymsigma2}
  \Sigma^2(L)=\chi L+C+O(L^{-1}) \qquad
  (L\rightarrow\infty)\;,
\end{equation}
where $\chi=1+2\alpha_0$ and $C=2\alpha_1$ (for the determination of
the constant term we have assumed that $R_2(s)$ falls off faster than
$1/s^2$, which is true for the cases considered in this article).
\subsection{Nearest-neighbor interaction ($k=1$)}
We begin with the simplest case, where the interaction is
restricted to nearest neighbors, so that the expression for the
partition function eq.~(\ref{ZN}) simplifies to
\begin{equation}
  \label{ZN1}
  Z_N(L)=\int\limits_0^{\infty}{\rm d}\xi_1\ldots
  \int\limits_0^{\infty}{\rm d}\xi_N\;
  \delta\Big(L-\sum_{i=1}^N\xi_i\Big)\prod_{j=1}^Nf(\xi_j)\;.
\end{equation}
By Laplace transformation with respect to $L$ we obtain 
\begin{eqnarray}
  g_N(t) &:=& \int\limits_0^{\infty}{\rm d}L\;Z_N(L)\,e^{-tL}
  \nonumber\\ &=& \bigg( \int\limits_0^{\infty}{\rm d}x\,
  f(x)\,e^{-tx}\bigg)^N =: \Big[g(t)\Big]^N\;.
\end{eqnarray}
The large $N$ limit of the partition function can now be calculated by
performing the Laplace inversion in saddle point approximation. 
We have 
\begin{eqnarray}
  \label{InvLaplace}
  Z_N(L) &=& \frac{1}{2\pi i}\int\limits_{c-i\infty}^{c+i\infty}
  {\rm d}t\;g_N(t)\,e^{Lt} \nonumber\\ &=& \frac{1}{2\pi i}
  \int\limits_{c-i\infty}^{c+i\infty} {\rm d}t\;
  e^{N(t\Delta+\log g(t))} \nonumber\\ &\sim& \Big[g(c)\Big]^N
  \,e^{Lc}\;,  
\end{eqnarray}
where $\Delta=L/N$ is the mean level spacing and $c$ is determined 
from the saddle point equation 
\begin{equation}
  \label{saddlepoint}
  \Delta+\frac{g'(c)}{g(c)}=0\;.
\end{equation}

The expression for the joint probability distribution of $n$
consecutive spacings eq.~(\ref{jointspace}) reduces in the present case
to 
\begin{displaymath}
  p(\xi_1,\ldots,\xi_n)=\frac{Z_{N-n}(L-\sum_{i=1}^n\xi_i)}{Z_N(L)}
  \prod_{j=1}^nf(\xi_j)\;.
\end{displaymath}
Using eq.~(\ref{InvLaplace}) and assuming $n\ll N$ we obtain 
\begin{equation}
  \label{asymjoint}
  p(\xi_1,\ldots,\xi_n)=\prod_{j=1}^n\,\frac{1}{g(c)}\,e^{-c\xi_j}\,
  f(\xi_j)\;.
\end{equation}
Note that the factor $g(c)^{-1}$ assures the proper normalization,
\begin{displaymath}
  \int\limits_0^{\infty}{\rm d}\xi_n\;p(\xi_1,\ldots,\xi_n)=
  p(\xi_1,\ldots,\xi_{n-1})\;.
\end{displaymath}
With help of eqs.~(\ref{nthP}) and (\ref{asymjoint}) we find for the
Laplace transformation of the $n$-th nearest-neighbor spacing
distribution 
\begin{equation}
  \label{LaplacenthP}
  g(n,t) := \int\limits_0^\infty {\rm d}s\,e^{-ts}\,P(n,s)
   = \left[\frac{g(c+t/\Delta)}{g(c)}\right]^{n+1}\;.
\end{equation}
Using the relation (\ref{relR2Pn}) we calculate the Laplace transform
of the two-point correlation function, eq.~(\ref{LaplaceR2}), 
\begin{equation}
  \label{LaplceR2k1}
  g_2(t)=\frac{g(c+t/\Delta)}{g(c)-g(c+t/\Delta)}\;.
\end{equation}
The small-$t$ asymptotic behavior is 
\begin{eqnarray}
  g_2(t) &=& \frac{1}{t}-1-\frac{1}{2\Delta}\frac{g''(c)}{g'(c)}+
  \nonumber\\ & & +{}\frac{3\,g''(c)^2-2\,g'(c)\,g^{(3)}(c)} 
  {12\,g'(c)^2}\frac{t}{\Delta^2}+ O(t^2)\;,
\end{eqnarray}
where we have employed the saddle point equation (\ref{saddlepoint}). 

At this point we specialize to the random matrix interaction
potential, i.e. we substitute eq.~(\ref{frm}). Its Laplace
transform reads  
\begin{equation}
  \label{gt}
  g(t)=\frac{\Gamma(\beta+1)}{t^{\beta+1}}
\end{equation}
and the saddle point equation (\ref{saddlepoint}) yields
\begin{equation}
  c=\frac{\beta+1}{\Delta}\;.
\end{equation}
By Laplace inversion of eq.~(\ref{LaplacenthP}) we then obtain the
$n$-th nearest-neighbor spacing distribution
\begin{equation}
  \label{pnbeta}
  P(n,s)=\frac{(\beta +1)^{(n+1)(\beta+1)}}{\Gamma\left((\beta+1)
  (n+1)\right)}\, s^{\beta+n(\beta+1)}\,{\rm e}^{-(\beta+1)s}\;. 
\end{equation}
In particular, the nearest-neighbor spacing distribution for
$\beta=1,2,4$ is 
\begin{eqnarray}
  \label{p124}
  P(s) &=& 4s\,{\rm e}^{-2s}\qquad\qquad\, (\beta=1)\;,\nonumber\\
  P(s) &=& \frac{27}{2}s^2\,{\rm e}^{-3s}\qquad\quad
  (\beta=2)\;,\nonumber\\ 
  P(s) &=& \frac{3125}{24}s^4\,{\rm e}^{-5s}\qquad (\beta=4)\;.
\end{eqnarray}
The Laplace transform of the two-point correlation function
eq.~(\ref{LaplceR2k1}) becomes
\begin{equation}
  \label{g2}
  g_2(t)=\frac{1}{(1+\frac{t}{\beta+1})^{\beta+1}-1}\;,
\end{equation}
from which $R_2(s)$ can be computed by Laplace inversion.
\begin{eqnarray*}
  R_2(s) &=& 1-\exp(-4s) \hspace{3.6cm} (\beta=1)\\
  R_2(s) &=& 1-\Big(\cos(3\sqrt{3}\,s/2)+\\ & & \hspace{1.2cm} 
  \sqrt{3}\,\sin(3\sqrt{3}\,s/2)\Big)\,e^{-9s/2} \qquad(\beta=2)\\
  R_2(s) &=& e^{-5s} \sum_{k=0}^{4}\exp\Big(5s\,e^{2\pi ik/5}+ 
  2\pi ik/5\Big) \hspace{0.4cm} (\beta=4)
\end{eqnarray*}
With help of eqs.~(\ref{g2expansion}) and (\ref{asymsigma2}) and the
small $t$ expansion of eq.~(\ref{g2}) we find the 
large $L$ behavior of the number variance $\Sigma^2(L)\sim\chi L+C$
with   
\begin{displaymath}
  \chi=\frac{1}{\beta+1}\;,\qquad 
  C=\frac{\beta(\beta+2)}{6(1+\beta)^2}\;.
\end{displaymath}
The spacing distribution and the two-point correlation function of the 
simplest model with $k=\beta=1$ are very close to the corresponding 
distributions of a number of different dynamical systems, whose spectral 
statistics can be calculated only numerically (see sects.\ref{pseudo} 
and \ref{discussion}). In lack of a deeper understanding of the spectral 
statistics of these systems, the plasma model is valuable in that it 
provides simple fitting distributions to these numerical results. 
We propose to call the statistics derived from the plasma model with 
$k=\beta=1$ the {\em Semi-Poisson statistics}. This name was originally 
motivated from the fact that if one takes an ordered Poisson distributed 
sequence $\{x_n\}$, the nearest-neighbor spacing distribution of the new 
sequence $\{y_n\}$ with $y_n=(x_n+x_{n+1})/2$ coincides with 
eq.~(\ref{semifish}) (of course the other correlation functions of 
the sequence $\{y_n\}$ are  different from those of the plasma model).

It is interesting to note that again starting from the sequence $\{x_n\}$ 
and dropping every second level one obtains a new sequence with precisely 
the same statistical distribution as the plasma model with $k=\beta=1$ 
\cite{seligman99a}. More generally, retaining only every $(p+1)-$th level 
of the sequence $\{x_n\}$ leads to the same statistical distribution as 
the plasma model with $k=1$ and $\beta=p$ \cite{seligman99a}.
\subsection{Interaction between nearest and next-to-nearest
  neighbors ($k=2$)}
Next we consider the one-dimensional gas, where the interaction is
restricted to nearest and next-to-nearest neighbors. 
In this case the partition function takes the form
\begin{eqnarray*}
  Z_N(L) &=& \int\limits_0^{\infty}{\rm d}\xi_1\ldots
  \int\limits_0^{\infty}{\rm d}\xi_N\;  
  \delta\Big(L-\sum_{i=1}^N\xi_i\Big)\times\\ & & \qquad
  \prod_{j=1}^Nf(\xi_j)f(\xi_j+\xi_{j+1})\;.
\end{eqnarray*}
As in the previous section we first compute the Laplace transform of
the partition function, 
\begin{eqnarray*}
  g_N(t) &:=& \int\limits_0^{\infty}{\rm d}L\;Z_N(L)\,e^{-tL}
  \\ &=& \int\limits_0^{\infty}{\rm d}\xi_1\ldots
  \int\limits_0^{\infty}{\rm d}\xi_N\; 
  \prod_{j=1}^N e^{-t\xi_j} f(\xi_j)f(\xi_j+\xi_{j+1})\;.
\end{eqnarray*}
To evaluate this integral we introduce the transfer operator 
\cite{vanhove50a}
\begin{equation}
  K(\xi,\xi')=\sqrt{f(\xi)}\,e^{-t\xi/2} f(\xi+\xi') 
  \sqrt{f(\xi')}\,e^{-t\xi'/2}\;,
\end{equation}
so that $g_N(t)$ can be rewritten as 
\begin{eqnarray}
  \label{KN}
  g_N(t) &=& \mbox{tr}\,K^N \nonumber\\
  &\equiv& \int\limits_0^{\infty}{\rm d}\xi_1\ldots
           \int\limits_0^{\infty}{\rm d}\xi_N\; 
  K(\xi_1,\xi_2 )K(\xi_2,\xi_3 )\ldots \nonumber\\
  & & \hspace{2cm}\ldots\, K(\xi_{N-1},\xi_{N} )K(\xi_N,\xi_1)\;.
\end{eqnarray}
The operator $K(\xi,\xi')$ is real symmetric and therefore admits
the eigenbasis expansion
\begin{equation}
  \label{eigexpansion}
  K(\xi,\xi')=\sum_j \lambda_j\,\phi_j(\xi)\,\phi_j(\xi')
\end{equation}
with real eigenvalues $\lambda_j$ and eigenfunctions $\phi_j(\xi)$,
\begin{equation}
  \label{inteq}
  \int\limits_0^{\infty}{\rm d}\xi'\,K(\xi,\xi')\,
  \phi_j(\xi')=\lambda_j \phi_j(\xi)\;.
\end{equation} 
The eigenfunctions are normalized according to
\begin{displaymath}
  \int\limits_0^{\infty}{\rm d}\xi\,\phi_j(\xi) 
  \phi_{j'}(\xi)=\delta_{j,j'}
\end{displaymath}
and we choose the ordering of the eigenvalues by decreasing magnitude
($\lambda_0>\lambda_1>\ldots$).  
Consequently, in the large $N$ limit the Laplace transform of the
partition function, eq.~(\ref{KN}), reduces to
\begin{equation}
  g_N(t)=\Big[\lambda_0(t)\Big]^N\;, 
\end{equation}
where we have explicitly indicated the $t$-dependence of the
eigenvalue. 
Again we perform the Laplace inversion in saddle point approximation,
which results in $t$ being fixed to $t=c$, 
\begin{equation}
  \label{ZN2}
  Z_N(L)\sim \Big[\lambda_0(c)\Big]^N\,e^{Lc}\;,
\end{equation}
where $c$ is determined from the saddle point equation
\begin{equation}
  \label{saddlepoint2}
  \Delta+\frac{\lambda_0'(c)}{\lambda_0(c)}=0\;.  
\end{equation}
Next we calculate the joint probability distribution of $n$
consecutive spacings, eq.~(\ref{jointspace}). 
It turns out to be convenient to write the eigenfunctions in the form 
\begin{equation}
  \label{eigfctform}
  \phi(t;\xi)=\sqrt{f(\xi)}\,e^{-t\xi/2}\,\psi(t;\xi)
\end{equation}
(again the dependence on $t$ is explicitly indicated).
We express the Laplace transform of the integral on the right hand side of
eq.~(\ref{jointspace}) in terms of the transfer operator $K$, use the
eigenfunction expansion eq.~(\ref{eigexpansion}), perform the Laplace
inversion in saddle point approximation (the saddle point equation is
identical with eq.~(\ref{saddlepoint2})), and arrive at
\begin{eqnarray}
  \label{jointspace2}
  p(\xi_1,\ldots,\xi_n) &=& \frac{1}{\lambda_0(c)^{n-1}}\, 
  \psi_0(c;\xi_1)\,\psi_0(c;\xi_n)\,e^{-c\sum_{i=1}^n\xi_i}
  \nonumber\\ & & f(\xi_n) 
  \prod_{j=1}^{n-1} f(\xi_j)f(\xi_{j}+\xi_{j+1})\;.
\end{eqnarray}
With the help of eq.~(\ref{jointspace2}) we can now calculate all 
correlation functions. The simplest is the nearest-neighbor spacing 
distribution 
\begin{equation}
  \label{spacingk2}
  P(s)=\Delta\,\Big[\phi_0(c;s\Delta)\Big]^2\;.
\end{equation}
To calculate the asymptotic behavior of the number variance we first 
take the Laplace transform of the $n$-th nearest-neighbor spacing 
distribution eq.~(\ref{nthP}) which yields
\begin{eqnarray}
  \label{Lnthspacing2}
  g(n,t) &=& \sum_j\left(\frac{\lambda_j(c+t/\Delta)}
  {\lambda_0(c)}\right)^n \times \nonumber\\ & &
  \bigg[\int\limits_0^{\infty}{\rm d}\xi\,\phi_0(c;\xi)\,
  \phi_j(c+t/\Delta;\xi)\,e^{-t\xi/2\Delta}\bigg]^2\;.
\end{eqnarray}
Hence, by eq.~(\ref{relR2Pn}), the Laplace transform of the two-level 
correlation function takes the form 
\begin{eqnarray}
  \label{Ltwolevel2}
  g_2(t) &=& \sum_j\frac{\lambda_0(c)}{\lambda_0(c)-
  \lambda_j(c+t/\Delta)} \times \nonumber\\ & &
  \bigg[\int\limits_0^{\infty}{\rm d}\xi\,\phi_0(c;\xi)\,
  \phi_j(c+t/\Delta;\xi)\,e^{-t\xi/2\Delta}\bigg]^2\;.
\end{eqnarray}
For small $t$ this becomes
\begin{equation}
  \label{g2asymt}
  g_2(t)=\frac{1}{t}-\frac{1}{2\Delta}
  \frac{\lambda_0''(c)}{\lambda_0'(c)}-1+O(t)\;,
\end{equation}
where we have used the saddle point equation (\ref{saddlepoint2}) and 
\begin{displaymath}
  \frac{1}{\Delta}\int\limits_0^\infty{\rm d}\xi\,
  \xi\,\phi_0(c,\xi)^2=1\;, 
\end{displaymath}
which follows from eq.~(\ref{spacingk2}) and the definition of the mean 
level spacing $\Delta$ (implying $\int{\rm d}s\,sP(s)=1$).

We now specialize to the random matrix interaction eq.~(\ref{frm}) 
and restrict ourselves to the case of $\beta$ integer. 
The integral equation (\ref{inteq}) with the eigenfunctions in the 
form of eq.~(\ref{eigfctform}) then reads
\begin{equation}
  \label{inteq2} 
  \int\limits_0^{\infty}{\rm d}\xi'\,e^{-t\xi'}\xi'^{\beta}
  (\xi'+\xi)^{\beta}\psi_j(t;\xi')=\lambda_j(t)\psi_j(t;\xi)\;.
\end{equation}
It is clear from eq.~(\ref{inteq2}) that the $t$ dependence
factorizes and  
\begin{eqnarray}
  \label{sc}
  \lambda_j(t) &=& t^{-2\beta-1}\lambda_j(1)\,,\nonumber\\
  \psi_j(t;\xi) &=& t^{(\beta+1)/2}\psi(1;t\xi)\;.
\end{eqnarray}
This permits us to determine $c$ from the saddle point 
equation (\ref{saddlepoint2}), which yields
\begin{equation}
  \label{spvalue}
  c=\frac{2\beta+1}{\Delta}\;.
\end{equation}
Using eqs.~(\ref{g2expansion}), (\ref{asymsigma2}), and 
(\ref{g2asymt}) we then find the asymptotic behavior of the 
number variance,
\begin{displaymath}
  \Sigma^2(L)\sim \chi L\;,\qquad \chi=\frac{1}{2\beta+1}\;.
\end{displaymath}
In order to determine the spacing distributions we actually need to 
solve the integral equation (\ref{inteq2}). It is straightforward to 
see that it has $\beta+1$ solutions each being a polynomial 
of degree $\beta$,
\begin{equation}
  \label{inteqsol}
  \psi(t;\xi)=\sum_{j=0}^{\beta}a_j\,\xi^{j}\;.
\end{equation}
The coefficients $a_j$ are easily obtained by substituting 
eq.~(\ref{inteqsol}) into eq.~(\ref{inteq2}). The spacing 
distributions then follow from eq.~(\ref{jointspace2}) and  
eq.~(\ref{nthP}). Table~\ref{table2} shows the explicit expression 
for the nearest and next-to-nearest-neighbor spacing 
distribution for $\beta$=1, 2, and 4.
\subsection{General case}
Finally we extend the calculation of the previous 
section to an interaction between an arbitrary number $k$ of 
neighboring particles. In this general case the Laplace 
transform of the partition function (\ref{ZN}) takes the form
\begin{eqnarray}
  \label{gNh}
  g_N(t) &=& \int\limits_0^{\infty}{\rm d}\xi_1\ldots
  \int\limits_0^{\infty}{\rm d}\xi_N\; 
  \prod_{j=1}^N e^{-t\xi_j} \times \nonumber\\
  & & f(\xi_j)f(\xi_j+\xi_{j+1})\cdots 
  f(\xi_j+\ldots+\xi_{j+k-1})\;.
\end{eqnarray}
Again we seek to express $g_N(t)$ as the trace of the $N$-th power 
of a transfer operator. To this end we define \cite{vanhove50a}
%\widetext
%\top{-2.8cm}
\begin{eqnarray}
&&K({\bf x},{\bf x'})= \delta(x_2-x_1')\,\delta(x_3-x_2')
  \cdots \delta(x_{k-1}-x'_{k-2}) \times \nonumber\\
&& e^{-tx_1/2}\sqrt{f(x_1)}\sqrt{f(x_1+x_2)}\cdots
  \sqrt{f(x_1+\ldots+x_{k-1})} \times\nonumber\\ & &
  f(x_1+\ldots+x_{k-1}+x'_{k-1})\, 
  \sqrt{f(x'_1+\ldots+x'_{k-1})}\cdots\times \nonumber\\
&&  \sqrt{f(x'_{k-2}+x'_{k-1})}\sqrt{f(x'_{k-1})}\,
  e^{-tx'_{k-1}/2}\;,
\label{transferop}  
\end{eqnarray}
%\narrowtext
where ${\bf x}=(x_1,\ldots,x_{k-1})$ and ${\bf x'}$ is defined 
likewise. One then easily verifies that
\begin{eqnarray*}
  g_N(t) &=& \mbox{tr}\, K^N \\
  &\equiv& \int{\rm d}{\bf x}_1\ldots\int{\rm d}{\bf x}_N\, 
  K({\bf x}_1,{\bf x}_2)\times\\ & & \qquad\qquad\qquad
  K({\bf x}_2,{\bf x}_3)\ldots K({\bf x}_N,{\bf x}_1)\;.
\end{eqnarray*}
It is clear from the definition (\ref{transferop}) that the transfer 
operator obeys the symmetry relation
\begin{equation}
  K({\bf x},{\bf x'})=K({\bf x'}^T,{\bf x}^T)\;,
\end{equation}
where we use the notation ${\bf x}^T=(x_{k-1},\ldots,x_1)$. The 
presence of this symmetry permits the eigenbasis expansion 
\begin{equation}
  K({\bf x},{\bf x'})=\sum_j \lambda_j\,\phi_j({\bf x})\,
  \phi_{j}({\bf x'}^T)\;,
\end{equation}
with the eigenvalues $\lambda_j$ and eigenfunctions $\phi_j({\bf x})$ 
obeying 
\begin{equation}
  \label{eigprobgen}
  \int{\rm d}{\bf x'}\,K({\bf x},{\bf x'})\,\phi_j({\bf x'})=\lambda_j
  \phi_j({\bf x})
\end{equation}
and the normalization
\begin{equation}
  \label{normalization}
  \int{\rm d}{\bf x}\,\phi_j({\bf x})\,\phi_{j'}({\bf x}^T)=
  \delta_{j j'}\;.
\end{equation}
The above considerations show that in the limit of large $N$ the 
Laplace transform of the partition function, eq.~(\ref{gNh}), can 
again be expressed as
\begin{equation}
  g_N(t)=\Big[\lambda_0(t)\Big]^N\;,
\end{equation}
where $\lambda_0$ is the largest eigenvalue of the transfer operator
(\ref{transferop}) which parametrically depends on $t$. 
Hence the form of the partition function and the saddle point 
equation will also be the same as in the last section, see 
eqs.~(\ref{ZN2}) and (\ref{saddlepoint2}).

In the following it will be useful to introduce the functions 
$\psi_j(t;{\bf x})$ by writing $\phi_j(t;{\bf x})$ in the form 
(again we indicate the $t$-dependence explicitly)
\begin{equation}
  \phi_j(t;{\bf x})=\psi_j(t;{\bf x})\,\sqrt{R(t;{\bf x})}\;,
\end{equation}
where
\begin{eqnarray}
  R(t;{\bf x}) &=& \exp\Big(-t\sum_{j=1}^{k-1}x_j\Big)\times\nonumber\\
  & & \qquad\quad\prod_{j=0}^{k-2}\,\prod_{i=1}^{k-j-1} 
  f(x_i+\ldots+x_{i+j})\;.
\end{eqnarray}
From eq.~(\ref{eigprobgen}) we obtain the equation
\begin{eqnarray}
  \label{eqforpsigen}
  &\int\limits_{0}^{\infty}{\rm d}\xi_k& \, e^{-t\xi_k} f(\xi_k)\,
  f(\xi_{k}+\xi_{k-1})\ldots f(\xi_k+\ldots+\xi_1)\times\nonumber\\
  & & \psi_j(t;\xi_{2},\ldots,\xi_{k}) =
  \lambda_j(t)\,\psi_j(t;\xi_{1},\ldots,\xi_{k-1})\;,
\end{eqnarray}
which in conjunction with the normalization condition 
\begin{equation}
  \int\limits_0^{\infty}{\rm d}{\bf x}\,R(t;{\bf x})\,\psi_j(t;{\bf x})\,
  \psi_{j'}(t;{\bf x}^T)=\delta_{j j'}
\end{equation}
determines $\psi_j(t;{\bf x})$ and $\lambda_j(t)$.

For the calculation of the joint probability distribution of $n$ 
consecutive spacings, eq.~(\ref{jointspace}), we follow the same 
steps as in the last section, except that we have to distinguish 
two cases, namely $n<k-1$ and $n\ge k-1$. In the former case,
\begin{eqnarray}
  \label{pcase1}
  p(\xi_1,&\ldots&,\xi_n)=\int\limits_0^\infty{\rm d}\xi_{n+1}
  \ldots\int\limits_0^\infty{\rm d}\xi_{k-1}\nonumber\\ & &
  \phi_0(c;\xi_1,\ldots,\xi_{k-1})\,\phi_0(c;\xi_{k-1},\ldots,\xi_1)\;,
\end{eqnarray}
where $c$ is determined from the saddle point equation 
(\ref{saddlepoint2}), while in the latter case,
\begin{eqnarray}
  \label{pcase2}
  p(&\xi_1&,\ldots,\xi_n)= \lambda_0^{-n+k-1} 
  \psi_0(c;\xi_n,\ldots,\xi_{n-k+2})\,
  e^{-c\sum\limits_{i=1}^n\xi_i}\nonumber\\ & &
  \bigg(\prod_{j=0}^{k-1}\prod_{i=1}^{n-j}
  f(\xi_i,\ldots,\xi_{i+j})\bigg)\,\psi_0(c;\xi_1,\ldots,\xi_{k-1})\;.
\end{eqnarray}
Eqs.~(\ref{pcase1}) and (\ref{pcase2}) in conjunction with 
eq.~(\ref{nthP}) allow the calculation of the spacing distributions.
For example, the nearest-neighbor spacing distribution becomes
\begin{eqnarray}
  \label{pofsgen}
  P(s)&=&\int\limits_0^\infty{\rm d}\xi_1\ldots
  \int\limits_0^\infty{\rm d}\xi_{k-1}\,\delta(s-\xi_j/\Delta)
  \nonumber\\ & & \quad\phi_0(c;\xi_1,\ldots,\xi_{k-1})\,
  \phi_0(c;\xi_{k-1},\ldots,\xi_1)\;,
\end{eqnarray}
where $j$ may take any value from 1 to k-1. In order to determine the 
asymptotic behavior of the number variance, we follow the same procedure 
as in the last section, i.e. we calculate the Laplace 
transform $g(n,t)$ of the $n$-th nearest-neighbor spacing distribution, 
sum this over $n$ to obtain the Laplace transform of the two-point 
correlation function, see eq.~(\ref{relR2Pn}), and find its small-$t$ 
asymptotic, which determines the asymptotic behavior of the number 
variance, see eq.~(\ref{asymsigma2}).
For $n<k-2$ we get
\begin{eqnarray*}
  &g&(n,t)=\int\limits_0^\infty{\rm d}\xi_1\ldots
  \int\limits_0^\infty{\rm d}\xi_{n+1}\,
  e^{-\frac{t}{\Delta}\sum\limits_{i=1}^{n+1}\xi_i}
  \int\limits_0^\infty{\rm d}\xi_{n+2}\ldots\\ & & \ldots
  \int\limits_0^\infty{\rm d}\xi_{k-1}\,
  \phi_0(c;\xi_1,\ldots,\xi_{k-1})\,\phi_0(c;\xi_{k-1},\ldots,\xi_1)\;,
\end{eqnarray*}
whose behavior for small $t$ is simply $g(n,t)=1+O(t)$ due to 
the normalization eq.~(\ref{normalization}). For $n\ge k-2$ we find
\begin{eqnarray*}
  &g&(n,t)=\sum_j\Big(\frac{\lambda_j(c+t/\Delta)}{\lambda_0(c)}
  \Big)^{n-k+2}\bigg[\int\limits_0^\infty{\rm d}\xi_1\ldots
  \int\limits_0^\infty{\rm d}\xi_{k-1}\\ & &
  e^{-\frac{t}{2\Delta}\sum\limits_{i=1}^{k-1}\xi_i}
  \phi_0(c;\xi_1,\ldots,\xi_{k-1})\,\phi_j(c+\frac{t}{\Delta};
  \xi_{k-1},\ldots,\xi_1)\bigg]^2.
\end{eqnarray*}
Summing over $n$ and expanding asymptotically for small $t$ we obtain 
\begin{displaymath}
  \sum_{n=k-2}^{\infty}g(n,t)=\frac{1}{t}-\frac{1}{2\Delta}\,
  \frac{\lambda_0''(c)}{\lambda_0'(c)}-(k-1) + O(t)\;,
\end{displaymath}
where we have used the saddle point equation (\ref{saddlepoint2}) and 
$\int{\rm d}s\,sP(s)=1$ with $P(s)$ given by eq.~(\ref{pofsgen}).
Since the $k-2$ remaining terms each contribute $1+O(t)$, we finally get
\begin{equation}
  \label{g2asymtgen}
  g_2(t)=\frac{1}{t}-\frac{1}{2\Delta}
  \frac{\lambda_0''(c)}{\lambda_0'(c)}-1+O(t)\;.
\end{equation}
Note that this expression depends on $k$ only by the largest 
eigenvalue of the corresponding transfer operator and it is 
identical with eq.~(\ref{g2asymt}).

At this point we again specialize to the random matrix interaction 
eq.~(\ref{frm}) with integer $\beta$. 
The integral equation (\ref{eqforpsigen}) which determines $\lambda_j(t)$ 
and $\psi_j(t;{\bf x})$ then reads
\begin{eqnarray}
  \label{inteqgen} 
  &\int\limits_{0}^{\infty}{\rm d}\xi_k& \,\xi_k^\beta\,
  (\xi_{k}+\xi_{k-1})^\beta\ldots(\xi_k+\ldots+\xi_1)^\beta e^{-t\xi_k} 
  \times\nonumber\\ & & \psi_j(t;\xi_{2},\ldots,\xi_{k}) =
  \lambda_j(t)\,\psi_j(t;\xi_{1},\ldots,\xi_{k-1})\;,
\end{eqnarray}
The $t$-dependence of the eigenvalues and eigenfunctions factorizes and  
\begin{eqnarray}
  \label{scgen}
  \lambda_j(t) &=& t^{-k\beta-1}\lambda_j(1)\,,\nonumber\\
  \psi_j(t;{\bf x}) &=& t^{(k-1)(k\beta+2)/4}\psi(1;t{\bf x})\;.
\end{eqnarray}
This permits us to determine $c$ from the saddle point 
equation (\ref{saddlepoint2}), which yields
\begin{equation}
  \label{spvaluegen}
  c=\frac{k\beta+1}{\Delta}\;.
\end{equation}
Using eqs.~(\ref{g2expansion}), (\ref{asymsigma2}), and 
(\ref{g2asymtgen}) we then find the asymptotic behavior of the 
number variance,
\begin{equation}
  \label{sigmaresult}
  \Sigma^2(L)\sim \chi L\;,\quad \chi=\frac{1}{k\beta+1}\qquad 
  (L\gg 1)\;.
\label{sigma2}
\end{equation}
In critical statistics this quantity is related with a certain 
(multi)fractal dimension
\cite{kravtsov}
\begin{equation}
\chi=\frac{\eta}{2d},
\end{equation}
where $d$ is the dimensionality of the system and $\eta=d-D_2$ and $D_2$ is
one of the multifractional exponent defined through the behavior of the
mean inverse participation ratio with the size of the system, $L$,  
\begin{equation}
\left< \int d^dx|\psi_n(x)|^4\right> \sim L^{-D_2} 
\end{equation}
In our model geometrical interpretation of non-zero values of 
$\chi$ (\ref{sigma2}) remains unclear.

In order to find the spacing distributions explicitly we need to 
determine $\psi_0$ and $\lambda_0$ from eq.~(\ref{inteqgen}) and 
then use eq.~(\ref{pcase1}) or (\ref{pcase2}) together with
eq.~(\ref{nthP}). The solutions of eq.~(\ref{inteqgen}) have the form
\begin{eqnarray}
  \label{psiformgen}
  \psi(t;\xi_1,\ldots,\xi_{k-1}) &=& \sum_{i_1=0}^{\beta}
  \sum_{i_2=0}^{3\beta}\ldots\sum_{i_{k-1}=0}^{(k-1)k/2}\nonumber\\
  & & \qquad a_{i_1i_2\ldots i_{k-1}}\,\xi_1^{i_1}\xi_2^{i_2}\ldots
  \xi_{k-1}^{i_{k-1}}\;.
\end{eqnarray}
The coefficients $a_{i_1i_2\ldots i_{k-1}}$ must be determined 
numerically by substituting eq.~(\ref{psiformgen}) into 
eq.~(\ref{inteqgen}). In tables \ref{table1}, \ref{table2}, and 
\ref{table3} the explicit form of $P(s)$ and $P(1,s)$ are shown for
$k$=1, 2, and 3 respectively, where in each case the distributions for 
$\beta$=1, 2, and 4 are given.

These calculations become more and more tedious as $k$ or 
$\beta$ increase. However it is straightforward to find the large 
$s$ asymptotic of the spacing distributions,
\begin{equation}
  \label{largespacing}
  P(n,s)\sim e^{-\Lambda s}\;,\quad \Lambda=k\beta+1 \qquad (s\gg 1)\;.
\end{equation}
Note that in our model the asymptotics of the spacing distributions and 
the number variance are related as follows, 
\begin{equation}
\Lambda=\frac{1}{\chi}.
\label{L}
\end{equation}
For critical distribution of of the Anderson model at MIT point it is often
assumed \cite{altshuler} that
\begin{equation}
\Lambda=\frac{1}{2\chi}.
\label{L2}
\end{equation}
To derive this relation it was assumed \cite{altshuler} that the
probability to find $n$ levels in an interval which contains in average $L$
levels has  the Gaussian form
\begin{equation}
P_n(L)\sim \exp (-\frac{(n-L)^2}{2\Sigma^2(L)}),
\label{gaussian}
\end{equation}
where $\Sigma^2(L)$ as above is the number variance.

Therefore the probability that there is no levels inside this interval (i.e. the
nearest-neighbor distribution) is
\begin{equation}
P_0(L)\sim \exp (-\frac{L^2}{2\Sigma^2(L)}).
\end{equation}
Assuming that, when $L\rightarrow \infty$, $\Sigma^2(L)\rightarrow \chi L$ one
gets Eq.~(\ref{L2}).

But it is clear that the assumption that $P_n(L)$ has the Gaussian form
(\ref{gaussian}) even for small $n$ is an oversimplification and cannot be
strictly valid in general. In our short-range plasma model $P_n(L)$ is like the one 
for the
Poissonian process which gives Eq.~(\ref{L}). In the Gaudin model with
$\beta=2$ there is no simple relation between $\Lambda$ and $\chi$ (see
Eqs~(\ref{chi}) and (\ref{lambda}) (though for small $\chi$ one obtains
Eq.~(\ref{L2})).

This difference between the Gaudin model when $\chi \rightarrow 0$ and the
relation (\ref{L}) derived in our short range plasma model is probably
related with the different truncation of the interaction between two
levels. In the former model the distance between two levels is important while in
the later one only the number of levels between the two given levels is taken into
account. 

Numerical calculations of the Anderson model at the MIT point and certain
analytical arguments \cite{nishigaki}, \cite{tsvelik} based on the kernel
(\ref{kernel}) are in the favor of the relation (\ref{L2}) at least in the
limit of small $\chi$. The short range plasma model demonstrates that there
exist (mathematical) models of intermediate statistics which do not obey 
this relation.

\section{Spectral statistics of pseudointegrable billiards}
\label{pseudo}

Pseudointegrable systems, as introduced by Richens and Berry 
\cite{richens81a}, are dynamical systems which are neither
integrable nor chaotic. The difference between integrable and
pseudointegrable systems is that for the former the phase space is
foliated into  2-dimensional  surfaces with genus $g=1$ (i.e. tori) while
for the latter it is foliated into surfaces of higher genus.
The simplest example of pseudointegrable systems is plane
polygonal billiards whose angles are all rational multiples
of $\pi$. For these models the genus of the corresponding surface 
is given by 
\begin{equation}
  \label{genus}
  g=1+\frac{N}{2}\,\sum_k\,\frac{m_k-1}{n_k}\;,
\end{equation}
where $m_k\pi/n_k$ are the interior vertex angles of the
polygon, $N$ is the least common multiple of the integers $n_k$, and
the sum is taken over all vertices of the polygon. The polygons with
$g=1$ (e.g. rectangle, equilateral triangle) are integrable, whereas
those with $g\ge 2$ are pseudointegrable.
In Refs.~\cite{shudo93a}, \cite{shudo94a} the spectral statistics
of a number of pseudointegrable polygonal billiards have been investigated 
numerically (and in the last reference even experimentally using a microcavity)
and it was found that they display level repulsion but 
deviate from random matrix theory.

In \cite{bogomolny99a} a number of different models had been 
considered numerically which clearly demonstrate the existence of
intermediate statistics. In this Section we present a more detailed
analysis of the data.

We consider the sequence of pseudointegrable billiards (with Dirichlet
boundary conditions) in the shape
of the right triangle with one angle equal  $\pi/n$, where
$n=5,7,8,9,\ldots,30$ 
(the triangles with $n=3,4,6$ are integrable) and using the 
boundary integral method we have calculated the first 20000-30000 levels 
for all these triangles (for some triangles even up to 80000 levels).  

\begin{figure}[t]
  \begin{center}
    \leavevmode
    \epsfig{figure=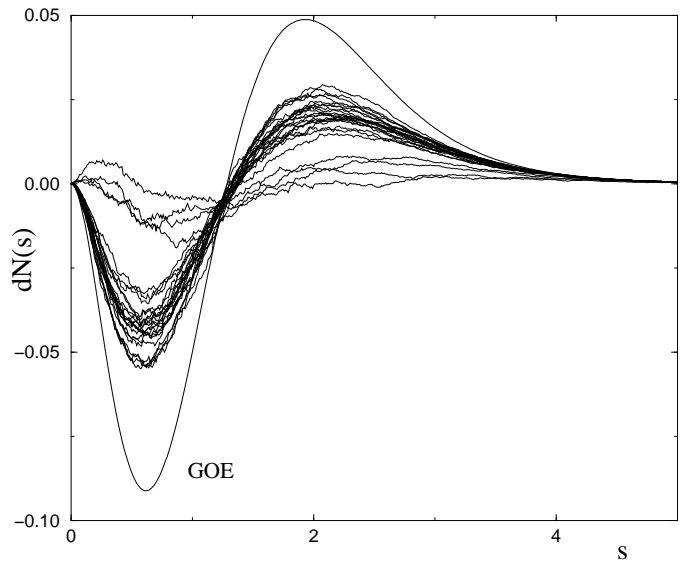,width=8.5cm,angle=270}
  \end{center}
  \caption{Difference between the cumulative spacing distribution for
    the pseudointegrable rational right triangles with $n\le30$
    (calculated from the levels 5000--20000) and the Semi-Poisson
    curve. See explanation in the text.}
  \label{fig:thirtytri}
\end{figure}

Figure~\ref{fig:thirtytri} shows the difference between the 
cumulative nearest-neighbor spacing distribution 
\begin{equation}
N(s)=\int_0^s{\rm d}x\,P(x)
\end{equation}
(calculated from the levels 5000--20000) for all triangles
and the Semi-Poisson prediction
\begin{equation}
N_1(s)=1-(2s+1)\exp(-2s)
\end{equation}
which is obtained from our model with $\beta=k=1$ (see Eq.~(\ref{p124})).

Roughly four close groups of curves are observed. For the first group, 
consisting of the curves corresponding to triangles with $n$=5, 8, 10, 12, 
this difference is less than one percent and consequently the spectral
statistics of these billiards is quite well described by the Semi-Poisson
model. The remaining 
three groups correspond to the triangles with $n$=7, 14, 18, with 
$n$=9, 16, 20, 24, 30, and the rest. 

The first conclusion from this and many others figures is that spectral
statistics of pseudointegrable systems is not universal and depends on the
billiard angles.  The grouping of the triangles which emerges from
Fig.~\ref{fig:thirtytri} does not agree with the classification of the
billiards according to their geometrical genus $g$. In first
approximation the spectral statistics of the above triangular billiards are
reasonably well described by a quantity $q$ which we proposed to 
call `arithmetical genus'
\begin{equation}
q(n)=\left \{ \begin{array}{cc} g(n), & n\;\mbox{odd}\\ 
\phi(n)/2,& n\;\mbox{even} \end{array} \right. .
\end{equation}
Hence $q$ is equal to the `geometrical genus' $g$ for $n$ odd but for even
$n$ it is given by half  the Euler function $\phi(n)$ (equal to
the number of integers not exceeding and relatively prime to $n$). The
first group now correspond to the triangles with $q=2$, the second to $q=3$,
the third to $q=4$, and the last to $q>4$. It seems that for $q>4$ the
spacing distribution does not change noticeably any more, but the resulting
distribution may differs from the Wigner-Dyson spacing distribution. We stress
that this classification is only an approximative one and  serves mostly for 
the crude arrangement of the spectral statistics of different triangles.

\begin{figure}[t]
  \begin{center}
    \leavevmode
    \epsfig{figure=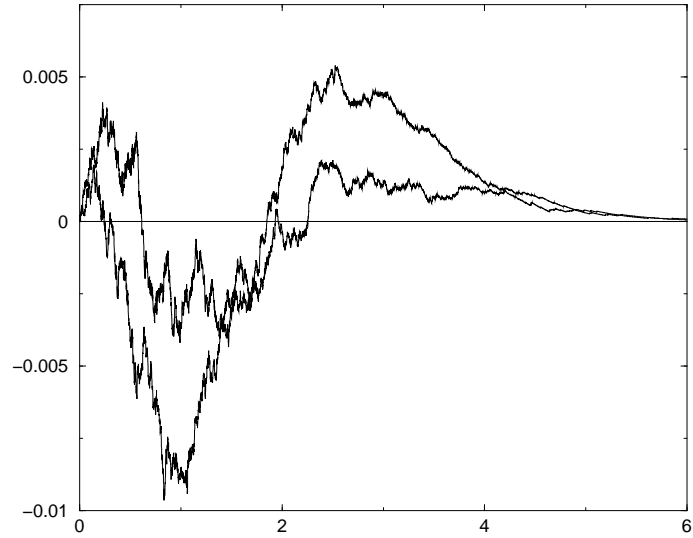,width=8.5cm,angle=270}
  \end{center}
  \caption{The same as in Fig.~\ref{fig:thirtytri} but for $n=12$. The thin
  line corresponds to 1-34000 levels and the thick one includes 34001-68000
  levels.} 
  \label{fig:ns}
\end{figure}

In Figure~\ref{fig:ns} we present the evolution of the
cumulative nearest-neighbor distribution for the triangular billiard with $n=12$
with increasing energy. It is clearly seen that for higher energy the
spacing distribution moves closer to the Semi-Poisson result (though a limiting
distribution may deviates from the Semi-Poisson one).  

\begin{figure}[t]
  \begin{center}
    \leavevmode
    \epsfig{figure=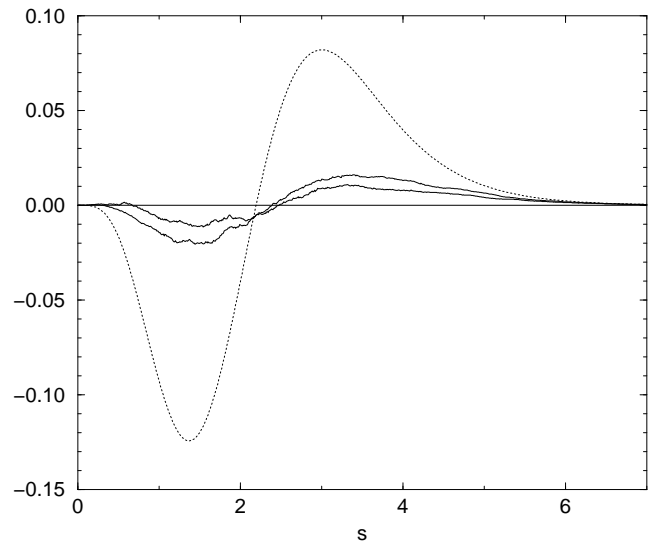,width=8.5cm}
  \end{center}
  \caption{The difference between the cumulative next-to-nearest spacing 
  distribution for $n=12$ and the Semi-Poisson prediction. Two curves are
  the same as in Fig.~\ref{fig:ns}. The dotted line corresponds to GOE}
  \label{fig:ns1}
\end{figure}

In Figure~\ref{fig:ns1} the difference between the cumulative next-to-nearest
distribution for the same triangle and the Semi-Poisson prediction (see
Table~\ref{table1}) is plotted. Once more we observed that this distribution 
is close to the Semi-Poisson result and that, with increasing energy, the 
agreement is better.

The main conclusion of this Section is that for certain pseudointegrable
systems the short-range spectral statistics is very close to the
Semi-Poisson statistics (though theoretical explication of this fact is not
yet clear).

\section{Summary and discussion}
\label{discussion}
In order to model intermediate spectral statistics, we considered a 
one-dimensional gas of energy levels interacting via a logarithmic 
pair potential, whose action is restricted to a small number of nearest 
neighbors. As shown in section~\ref{model}, its 
correlation functions can be calculated analytically, so the model 
deserves interest as a simple reference point for comparisons with numerics.
In section~\ref{pseudo} we performed a comparison with the spectral 
statistics of a number of pseudointegrable billiards and demonstrated that 
the plasma model with nearest-neighbor interaction (the Semi-Poisson model)
gives an excellent 
phenomenological description of the short-range spectral observables of
certain  pseudointegrable systems.
Unfortunately a full theoretical understanding of the spectral statistics 
of pseudointegrable systems is still lacking.

Of course there are many ways to interpolate between Wigner-Dyson and 
Poisson statistics. For the spacing distribution a well-known 
interpolation is the Brody distribution \cite{brody73a}, which shows 
fractional level repulsion. Since for intermediate spectral 
statistics $P(s)\sim s^\beta$ for small spacings, the Brody 
distribution cannot give an adequate description. 
A very natural way to construct interpolating ensembles is to take 
a weighted average between the ensemble of diagonal random matrices 
and one of the standard random matrix ensembles, e.g. for $\beta=1$ 
\begin{equation}
  \label{PoissonplusGOE}
  H=H_{\rm Poisson}+\lambda\, H_{\rm GOE}
\end{equation}
(see e.g. ref.~\cite{guhr89a}). However, as we checked numerically, 
for no value of $\lambda$  the resulting distribution is close to the 
distribution (\ref{semifish}), which we have found to be an excellent fit 
to the spacing distribution of the pseudointegrable triangles with $q=2$ 
and also to that of several other systems \cite{bogomolny99a}. Hence  
this interpolation do not seem  to be suitable for the description of 
intermediate statistics.

All data suggest that  a necessary requirement of theoretical
description of intermediate statistics is the screening of two-body
potential. Its precise form is not yet known (but see
\cite{kravtsov95}-\cite{tsvelik}) and may depend on the  problem  considered. 
\section{Acknowledgments}

The authors is greatly indebted to A. Pandey who had investigated the
short-range plasma model in the framework of band random matrices and whose
unpublished notes were useful to check our calculations.

It is a pleasure to thank O.~Bohigas, D.~Delande, A. Mirlin and 
G.~Montambaux for useful discussions. 
UG acknowledges a HSP2-fellowship of the 'Deutscher Akademischer 
Austauschdienst', and thanks the IPN Orsay for kind hospitality during 
an extended stay.

\pagebreak

%\widetext

\pagebreak

\pagebreak

\begin{table}
 \caption{The spacing distributions $P(s)$ and $P(1,s)$ for the 
          one-dimensional gas model with nearest-neighbor interaction.}
 \label{table1}
 \begin{tabular}{lccc}
  $(k=1)$ & $\beta=1$ & $\beta=2$ & $\beta=4$ \\ \hline 
  $P(s)$ & $4s\,e^{-2s}$ & $27/2\,s^2e^{-3s}$ &
  $3125/24\,s^4e^{-5s}$ \\ 
  $P(1,s)$ & $8/3\,s^3e^{-2s}$ & $243/40\,s^5e^{-3s}$ &
  $1953125/72576\,s^9e^{-5s}$ 
 \end{tabular}
\end{table}

\begin{table} 
 \caption{Same as Tab.~\ref{table1}, but with nearest- and 
          next-to-nearest-neighbor interaction.}
 \label{table2}
 \begin{tabular}{lccc}
  $(k=2)$ & $\beta=1$ & $\beta=2$ & $\beta=4$ \\ \hline 
  & $c_0=2.4773$ & $c_0=6.2603$ & 
                        $c_0=35.6018\,,\;c_1=252.5852$ \\ 
  $P(s)=s^\beta e^{-(2\beta+1)s}\sum_jc_js^j$ & 
         $c_1=6.0681$ & $c_1=24.9958$ & 
                        $c_2=826.3343\,,\;c_3=1630.9839$ \\
  & $c_2=3.7159$ & $c_2=41.1181$ & 
                        $c_4=2127.3725\,,\;c_5=1880.5911$ \\  
  & & $c_3=32.2768$ & $c_6=1103.6323\,,\;c_7=395.1819$ \\
  & & $c_4=10.4386$ & $c_8=66.6078$ \\
  & $d_0=2.5054$ & $d_0 = 5.9771$ & 
                        $d_0=29.4495\,,\;d_1=104.4681$ \\
  $P(1,s)=s^{3\beta+1} e^{-(2\beta+1)s}\sum_jd_js^j$ & 
      $d_1=3.068$ & $d_1=11.9326$ & 
                        $d_2=169.5742\,,\;d_3=164.1739$ \\
  & $d_2=0.7516$ & $d_2=9.5151$ & 
                        $d_4=103.3668\,,\;d_5=43.2920$ \\
  & & $d_3=3.3018$ & $d_6=11.9204\,,\;d_7=2.0002$ \\
  & & $d_4 =0.4746$ & $d_8=0.1586$ \\
 \end{tabular}

\end{table}
%\narrowtext
%
%\widetext
\begin{table}
\caption{Same as Tab.~\ref{table1}, but with three interacting 
          neighbors.}
 \label{table3}
 \begin{tabular}{lccc}
  $(k=3)$ & $\beta=1$ & $\beta=2$ & $\beta=4$ \\ \hline 
  & $c_0=2.1342$ & $c_0 = 4.9711$ 
                           & $c_0=24.0711\,,\;c_1=282.9984$ \\
  $P(s)=s^\beta e^{-(3\beta+1)s}\sum_jc_js^j$ & $c_1=7.8196$ 
        & $c_1=31.6156$ & $c_2=1615.27\,,\;c_3=5952.78$ \\
  & $c_2=11.6945$ & $c_2=93.9581$ 
                           & $c_4=15890.11\,,\;c_5=32666.46$ \\
  & $c_3=9.0743$ & $c_3=171.4838$ 
                           & $c_6=53678.64\,,\;c_7=72222.20$ \\
  & $c_4=3.6855$ & $c_4=212.8904$ 
                           & $c_8=80836.91\,,\;c_9=76044.48$ \\
  & $c_5=0.6259$ & $c_5=188.1012$ 
                      & $c_{10}=60480.73\,,\;c_{11}=40759.87$ \\
  & & $c_6=120.0046$ & $c_{12}=23247.21\,,\;c_{13}=11165.36$ \\
  & & $c_7=54.6739$ & $c_{14}=4473.73\,,\;c_{15}=1472.98$ \\
  & & $c_8=17.0418$ & $c_{16}=389.3105\,,\;c_{17}=79.6241$ \\
  & & $c_9=3.2829$ & $c_{18}=11.8577\,,\;c_{19}=1.1466$ \\
  & & $c_{10}=0.2968$ & $c_{20}=0.0541$ \\
  & $d_0=1.1340$ & $d_0=1.6110$ 
                          & $d_0=2.9196\,,\;d_1=29.5956$ \\
  $P(1,s)=s^{3\beta+1} e^{-(3\beta+1)s}\sum_jd_js^j$ 
       & $d_1=3.6334$ & $d_1=8.8782$ 
       & $d_2=146.1401\,,\;d_3=467.9231$ \\
  & $d_2=5.0189$ & $d_2=23.0847$ & 
         $d_4=1090.9150\,,\;d_5=1971.2442$ \\
  & $d_3=3.7112$ & $d_3=37.4284$ & 
         $d_6=2868.9524\,,\;d_7=3449.8348$ \\
  & $d_4=1.5037$ & $d_4=42.0839$ & 
         $d_8=3488.0347\,,\;d_9=3001.6809$ \\
  & $d_5=0.3100$ & $d_5=34.4593$ & 
         $d_{10}=2216.98\,,\;d_{11} = 1412.97$ \\
  & $d_6=0.0259$ & $d_6=20.9804$ & 
         $d_{12} = 779.3162\,,\;d_{13} = 372.3224$ \\
  & & $d_7=9.5327$ & $d_{14}=153.8623\,,\;d_{15}=54.8050$ \\
  & & $d_8=3.2010$ & $d_{16}=16.7526\,,\;d_{17}=4.3271$ \\
  & & $d_9=0.7747$ & $d_{18}=0.94224\,,\;d_{19}=0.1687$ \\
  & & $d_{10}=0.1284$ & 
            $d_{20}=0.0241\,,\;d_{21}=2.76* 10^{-3}$ \\
  & & $d_{11}=0.0131$ & 
  $d_{22}=2.5* 10^{-4}\,,\;d_{23}=1.9* 10^{-4}$ \\
  & & $d_{12}=6.2* 10^{-4}$ & $d_{24}=6.2*  10^{-7}$ \\
 \end{tabular}
\end{table}

%\narrowtext
%
%\widetext
\end{document}